\documentclass[cits]{PoS}
\usepackage{graphicx,amssymb,epsfig,epsf}
\usepackage{amsmath}
\usepackage{axodraw}

\title{Gluino-mediated FCNCs\\ in the MSSM with large tan$\bf\beta$}

\ShortTitle{Gluino-mediated FCNCs in the MSSM with large tan$\bf\beta$}

\author{\speaker{Lars HOFER}\\
        Institut f\"ur Theoretische Teilchenphysik,
        Karlsruhe Institute of Technology,\\
        D-76128 Karlsruhe, Germany\\
        E-mail: \email{lhofer@particle.uni-karlsruhe.de}}

\author{Ulrich NIERSTE\\
        Institut f\"ur Theoretische Teilchenphysik,
        Karlsruhe Institute of Technology,\\
        D-76128 Karlsruhe, Germany\\
        E-mail: \email{nierste@particle.uni-karlsruhe.de}}

\author{Dominik SCHERER\\
        Institut f\"ur Theoretische Teilchenphysik,
        Karlsruhe Institute of Technology,\\
        D-76128 Karlsruhe, Germany\\
        E-mail: \email{dominik@particle.uni-karlsruhe.de}}

\abstract{We present results of our study of $\tan\beta$-enhanced loop corrections in the Minimal Supersymmetric Standard Model (MSSM) with Minimal Flavour Violation (MFV). We demonstrate that these corrections induce flavour changing neutral current (FCNC) gluino couplings which have a large impact on the Wilson coefficient $C_8$ of the chromomagnetic operator. To illustrate the phenomenological consequences of this gluino-squark contribution to $C_8$, we briefly discuss its effect on the mixing-induced CP asymmetry in the decay $B_d\to\phi K_S$.}

\FullConference{European Physical Society Europhysics Conference on High Energy Physics\\
                 July 16-22, 2009\\
                 Krakow, Poland}

\begin{document}

\section{Introduction}
The Minimal Supersymmetric Standard Model (MSSM) contains two Higgs doublets
$H_u$ and $H_d$ coupling to up-type and down-type quark fields, respectively. The neutral components of these Higgs doublets acquire vacuum expectation values (vevs) $v_u$ and $v_d$ with the sum $v_u^2+v_d^2$ being fixed to $v^2\approx
(174\,\,\mbox{GeV})^2$ and the ratio $\tan\beta\equiv v_u/v_d$ remaining as a free parameter. Large values of $\tan\beta$ ($\sim50$) are theoretically motivated by bottom-top Yukawa unification, which occurs in SO(10) GUT models with
minimal Yukawa sector, and phenomenologically preferred by the anomalous magnetic moment of
the muon \cite{Bennet:2006}. Since a large value of $\tan\beta$ corresponds to $v_d\ll v_u$, it leads to enhanced corrections in amplitudes where the tree-level contribution is suppressed by the small vev $v_d$ but the loop-correction involves $v_u$ instead. %In such cases the ratio of one-loop to tree-level contribution receives an enhancement-factor $\tan\beta$ which may lift the loop-suppression rendering the ratio of order $\mathcal{O}(1)$. 

These $\tan\beta$-enhanced loop-corrections lead to a plethora of phenomenological consequences: They modify the relation between the down-type Yukawa couplings $y_{d_i}$ and the quark masses $m_{d_i}$ \cite{YukMass}, give corrections to the elements of the CKM matrix \cite{CKMcorr} and induce FCNC couplings of the neutral Higgs bosons to down quarks \cite{EffHiggs}. Recently we found that also FCNC couplings of gluinos and neutralinos to down quarks are generated in this way \cite{HNS}. In this article we explain how this FCNC gluino-couplings arise and discuss the phenomenological consequences.
%In order not to spoil the perturbative expansion a special treatment is required for these $\tan\beta$-enhanced corrections %to resum them to all orders. There are two possible ways to deal with them. The first is to consider an effective theory %with the SUSY-particles and the Higgs-fields intgrated out \cite{decoup}. This approach is valid for $M_{Susy}\gg v,%\,M_{A^0,H^0,H^{\pm}}$, i.e. within the decoupling limit; it can be extended beyond using an iterative method %\cite{decoupit}. The second possibility is to perform a diagrammatic, analytic resummation in the full MSSM without %assuming any hierarchy between $M_{Susy}$, $M_{A^0,H^0,H^{\pm}}$ and $v$ \cite{CGNW,HNS}. Going beyond the decoupling limit %is desirable since on the one hand $M_{Susy}\sim v$ is natural and on the other hand $\tan\beta$-enhanced effects in %couplings involving SUSY-particles like gluinos and neutralinos cannot be studied in an effective theory with these %particles integrated out. In ref. \cite{HNS} we discuss  

\section{Flavour-changing gluino coupling in naive MFV at large tan$\bf\beta$}
At tree-level the bottom mass $m_b$ is generated by coupling the $b$-quark to the Higgs field $H_d$ and is thus proportional to the small vev $v_d$. For this reason self-energy amplitudes $\Sigma^{RL}_{bi}$ ($i=d,s$) can be $\tan\beta$-enhanced compared to $m_b$ if they involve $v_u$ instead of $v_d$. In this article we discuss the impact of these enhanced self-energies in the framework of naive MFV as defined in Ref. \cite{HNS}. For the analysis of the analogous effects in the general MSSM we refer to Ref. \cite{CN}. For definiteness we focus on $b\to s$ transitions and parameterise the corresponding self-energy, which is generated by chargino-squark-loops, as
\begin{equation}
   \Sigma^{RL}_{bs}\,\,=\,\,V_{tb}^*\,V_{ts}\,m_b\,{\epsilon}_{FC}\,\tan\beta.
   \label{eq:SelfPara}
\end{equation}

In naive MFV the quark-squark-gluino coupling is flavour-diagonal at tree-level. 
  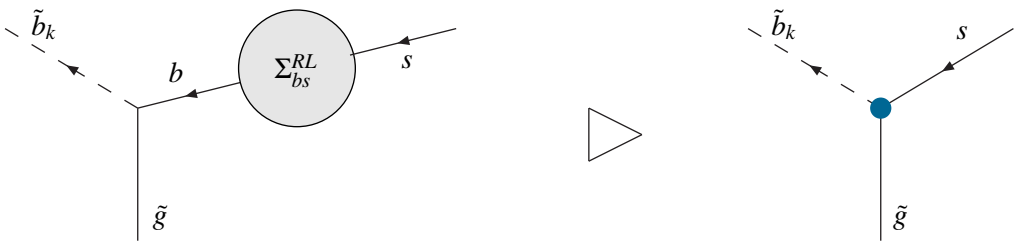
\begin{figure}[b]
     \begin{picture}(400,100)(0,-20)
        \SetWidth{0.5}
        \SetColor{Black}
        \SetWidth{0.5}
        \DashArrowLine(350,30)(300,60){5}
        \ArrowLine(400,60)(350,30)
        \Line(350,30)(350,-20)
        \Text(310,56)[lb]{$\tilde{b}_k$}
        \Text(380,56)[lb]{$s$}
        \Text(356,-16)[lb]{$\tilde{g}$}

        \Line(240,10)(240,30)
        \Line(240,10)(260,20)
        \Line(240,30)(260,20)

        \DashArrowLine(70,30)(20,60){5}
        \ArrowLine(110,40)(70,30)
        \GOval(130,45)(22,22)(0){0.9}
        \ArrowLine(190,60)(150,50)
        \Line(70,30)(70,-20)
        \Text(30,56)[lb]{$\tilde{b}_k$}
        \Text(170,45)[lb]{$s$}
        \Text(82,40)[lb]{$b$}
        \Text(76,-16)[lb]{$\tilde{g}$}
        \Text(122,38)[lb]{$\Sigma^{RL}_{bs}$}
        \SetColor{MidnightBlue}
        \Vertex(350,30){4}
     \end{picture}
\caption{FCNC gluino coupling for an on-shell $s$-quark induced by the $\tan\beta$-enhanced self-energy $\Sigma^{RL}_{bs}$}
\label{fig:FCNCgluino}
  \end{figure}
It receives flavour-changing loop-corrections among which we want to consider those induced by an insertion of the self-energy $\Sigma^{RL}_{bs}$ in the down-quark line (see fig.\,\ref{fig:FCNCgluino} for the case of an external $s$-quark). If the $s$-quark is on-shell, this correction is local and can be promoted to a FCNC gluino coupling. Since the $b$-quark propagator $- i/m_b$ ($m_s$ is set to zero) cancels the factor $m_b$ in $\Sigma^{RL}_{bs}$, the resulting coupling is proportional to 
\begin{equation}
   \kappa_{bs}\,\,=\,\,g_s\,V_{tb}^*\,V_{ts}\,{\epsilon}_{FC}\,\tan\beta.
   \label{eq:FCNCgluino}
\end{equation}
If $\tan\beta$ is large enough to compensate for the loop-factor ${\epsilon}_{FC}$, the coupling $\kappa_{bs}$ can be of order $\mathcal{O}(1)$ apart from the CKM factor $V_{tb}^*V_{ts}$, which preserves the MFV structure.

In order not to spoil the perturbative expansion a special treatment is required for these $\tan\beta$-enhanced corrections to resum them to all orders. This is usually done using an effective theory with the SUSY particles intgrated out keeping only Higgs fields and SM fields \cite{YukMass,CKMcorr,EffHiggs}. However, since we want to study $\tan\beta$-enhanced effects in the quark-squark-gluino-coupling, we cannot integrate out the gluino and squarks and this technique is not appropriate here. In Ref. \cite{HNS} instead the diagrammatic method developed in Ref. \cite{CGNW} is extended to the case of flavour-changing interactions. The result is that contributions of the form $(\textrm{loop}\times\tan\beta)^n$ can be included to all orders $n=1,2,...$ into the FCNC gluino coupling by replacing
\begin{equation}
   \epsilon_{FC}\,\tan\beta\,\,\longrightarrow\,\,\frac{\epsilon_{FC}\,\tan\beta}{1\,+\,(\epsilon_b\,-\,\epsilon_{FC})\,\tan\beta}
\end{equation}
in Eq. (\ref{eq:FCNCgluino}). Here $\epsilon_b$ denotes the counterpart of $\epsilon_{FC}$ in the parameterisation of the flavour-conserving self-energy $\Sigma^{RL}_b$ analogous to Eq. (\ref{eq:SelfPara}). Explicit formulae for $\epsilon_b$ and $\epsilon_{FC}$ can be found in Ref. \cite{HNS}.
%\begin{figure}[b]
% \begin{minipage}[t!]{0.45\linewidth}   
%     \begin{picture}(100,120)(300,-30)
%        \SetWidth{0.5}
%        \SetColor{Black}
%        \SetWidth{0.5}
%        \ArrowLine(300,60)(350,30)
%        \CArc(380,30)(30,180,0)
%        \DashArrowArcn(380,30)(30,180,0){3}
%        \ArrowLine(410,30)(460,60)
%        \Gluon(401,9)(440,-30){6}{3}
%        \Text(304,40)[lb]{$b$}
%        \Text(446,42)[lb]{$s$}
%        \Text(370,-15)[lb]{$\tilde{g}$}
%        \Text(370,64)[lb]{$\tilde{b}_k$}
%        \Text(410,-30)[lb]{$g$}
%        \SetColor{MidnightBlue}
%        \Vertex(410,30){4}
%     \end{picture}
%  \end{minipage}\hspace{0.1\linewidth}
%  \begin{minipage}[t!]{0.45\linewidth}
%        \includegraphics[width=\textwidth]{absc7c8g.eps}
%  \end{minipage}
%\caption{Left: Gluino diagram contributing to $C_8$. Right: Magnitudes of chargino and gluino contributions to $C_8$ %scanned over the MSSM parameter space. Right: }
%  \end{figure}

\section{Sizable effect in $C_8$}
The FCNC gluino coupling discussed in the last section gives rise to new contributions to the Wilson coefficients of the effective $\Delta B=1$ and $\Delta B=2$ Hamiltonians. Most of these contributions turn out to be numerically small for two reasons: Firstly, the FCNC gluino coupling is numerically small; for positive $\mu$ typical values are around $\kappa_{bs}\,\sim\, 0.1\,\cdot\,V_{tb}^*V_{ts}$. Secondly, unlike the higgsino-part of chargino diagrams, the gluino contributions suffer from a GIM suppression because the gluino coupling is universal for all quark flavours.  
\begin{figure}[b]
    \begin{minipage}[t]{0.35\linewidth}
        \includegraphics[width=\textwidth]{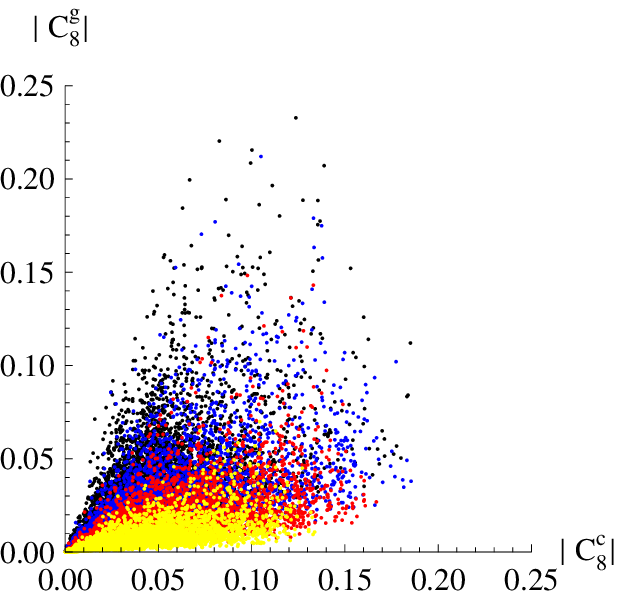}
    \end{minipage}\hspace{0.1\linewidth}
    \begin{minipage}[t]{0.55\linewidth}
         \includegraphics[width=\textwidth]{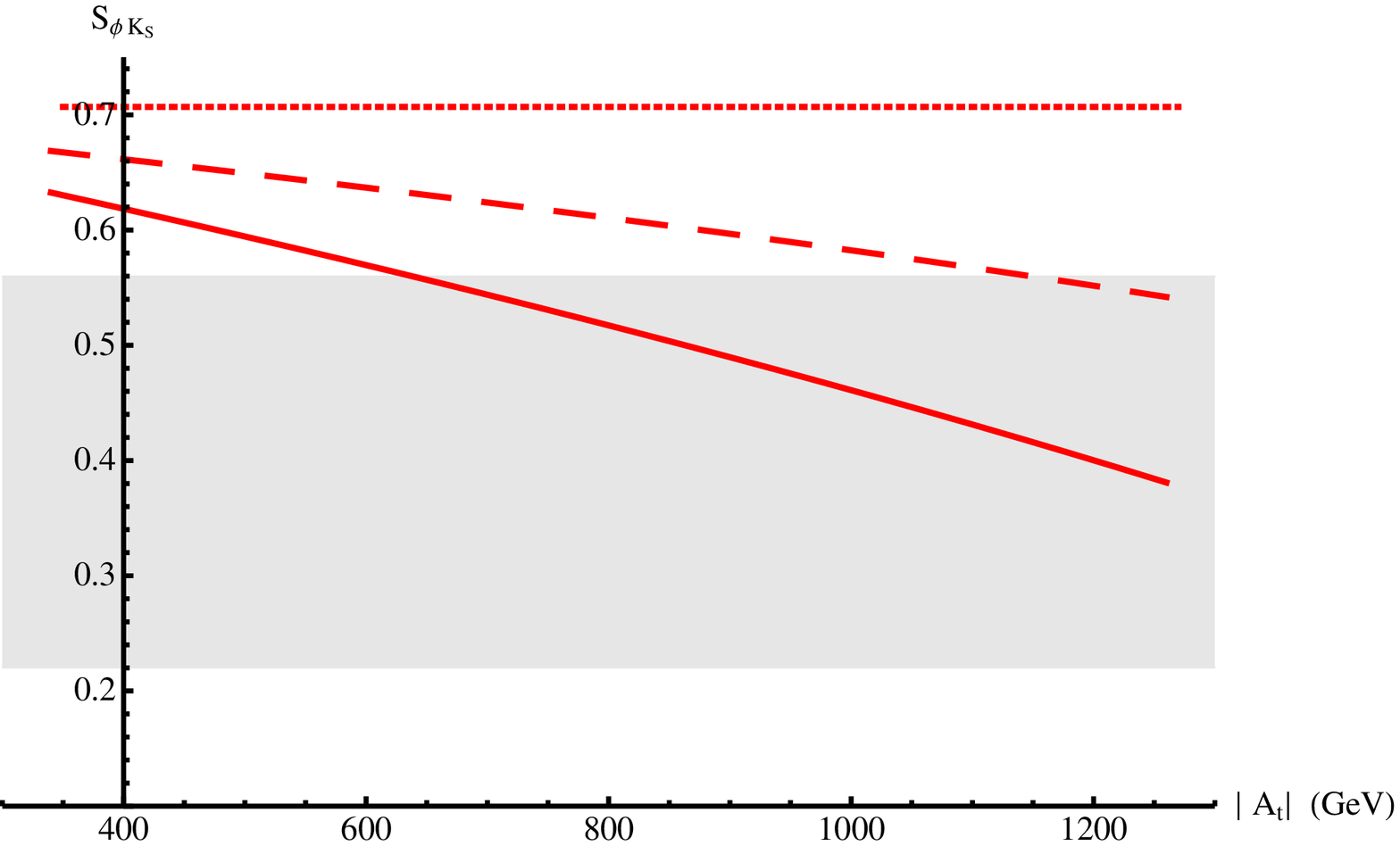}
    \end{minipage}
\caption{Left: Magnitudes of chargino and gluino contributions to $C_8$ scanned over the MSSM parameter space. Right: $S_{\phi K_S}$ as a funtion of $|A_t|$.}
\label{fig:plots}
\end{figure}

There is one exception: Chirally enhanced contributions to the magnetic and chromomagnetic operators $\mathcal{O}_7$ and $\mathcal{O}_8$ involve a left-right-flip in the squark-line which is proportional to the corresponding quark mass und thus distinguishes between different squark flavours. Whereas the corresponding contribution from gluino-squark-loops to $C_7$ is accidentally small, the one to $C_8$ can indeed contribute as much as the well known chargino-squark diagram. This can be seen from the left diagram in fig. \ref{fig:plots} where the magnitudes of both contributions $|C_8^c|$ and $|C_8^g|$ are shown for a scan over the MSSM parameter space with positive $\mu$. The colour code (yellow: $200\,\textrm{GeV}<\mu<\,400\,\textrm{GeV}$, red: $400\,\textrm{GeV}<\mu<600\,\textrm{GeV}$,
blue: $600\,\textrm{GeV}<\mu<800\,\textrm{GeV}$, black: $800\,\textrm{GeV}<\mu<1000\,\textrm{GeV}$) reflects the fact that the importance of $C_8^g$ grows with $\mu$. All points in the plot are in agreement with the constraints from $\mathcal{B}(\bar{B}\to X_s\gamma)$ and the experimental lower bounds for the sparticle and lightest Higgs Boson masses. Note that we allow for an arbitrary phase for the parameter $A_t$. However, to avoid the possibility of fulfilling the $\mathcal{B}(\bar{B}\to X_s\gamma)$ constraint by an unnatural fine-tuning of this phase, the additional condition $|C_7^{Susy}|<|C_7^{SM}|$ is imposed. 

As a consequence the $\tan\beta$-enhanced FCNC gluino coupling should affect those low energy observables with a strong dependence on $C_8$. To illustrate this fact we have plotted the mixing-induced CP asymmetry $S_{\phi K_S}$ of the decay $\bar{B}^0\to\phi\,K_s$ as a function of $|A_t|$ in the right diagram of fig.\,\ref{fig:plots}. The parameter point chosen for the plot fulfills all constraints mentioned above. The shaded area represents the experimental $1\sigma$ range, the dotted line the SM contribution in leading-order QCD factorisation. For the results corresponding to the dashed and the solid lines we have in addition taken into account the effects of $C_8^c$ and $C_8^c+C_8^g$, respectively. The plot demonstrates that for complex $A_t$ the gluino-squark contribution can indeed have a large impact on $S_{\phi K_S}$, especially if $|A_t|$ is large.
%\section*{Acknowledgements}


\begin{thebibliography}{99}
  \bibitem{Bennet:2006} \textbf{Muon G-2} Collaboration, G.W.~Bennet et. al., %\emph{Final report of the muon E821 anomalous %magnetic moment measurement at BNL}, 
         \emph{Phys. Rev. }\textbf{D73} (2006) 072003, [hep-ph/0602035].
  \bibitem{YukMass} L.J.~Hall, R.~Rattazzi, U.~Sarid, %\emph{The Top quark mass in supersymmetric SO(10) unification}, 
                   \emph{Phys. Rev.} \textbf{D50} (1994) 7048-7065, [hep-ph/9306309]; M.S.~Carena, M.~Olechowski, S.~Pokorski, C.E.M.~Wagner, %\emph{Electroweak symmetry breaking and bottom-top unification}, 
                   \emph{Nucl. Phys.} \textbf{B645} (2002) 155-187, [hep-ph/0207036].
  \bibitem{CKMcorr} T.~Blazek, S.~Raby, S.~Pokorski, %\emph{Finite supersymmetric threshold corrections to CKM matrix %elements in the large $\tan\beta$ regime}, 
                   \emph{Phys. Rev.} \textbf{D79} (2009) 035018, [0810.1613];
  \bibitem{EffHiggs} C.~Hamzaoui, M.~Pospelov, M.~Toharia, \emph{Phys. Rev. }\textbf{D59} (1999) 095005, [hep-ph/9807350];
                   K.S.~Babu, C.F.~Kolda, \emph{Phys. Rev. Lett. }\textbf{84} (2000) 228-231, [hep-ph/9909476];
                   A.J.~Buras, P.H.~Chankowski, J.~Rosiek, L.~Slawianowska, \emph{Nucl. Phys. }\textbf{B659} (2003) 3, [hep-ph/0210145].
  \bibitem{HNS} L.~Hofer, U.~Nierste, D.~Scherer, arXiv:0907.5408 [hep-ph].
  \bibitem{CN} A.~Crivellin, U.~Nierste, arXiv:0908.4404 [hep-ph]. 
  \bibitem{CGNW} M.~Carena, D.~Garcia, U.~Nierste, C.E.M.~Wagner, \emph{Nucl. Phys. }\textbf{B577} (2000) 88-120, [hep-ph/9912516].
\end{thebibliography}
\end{document}